# ASPECTOS FÍSICOS A CONSIDER AR EN LA CALIBRACIÓN RADIOMETRICA DE IMÁGENES SATELITALES

PHYSICAL ASPECTS TO CONSIDER IN RADIOMETRIC CALIBRATION OF SATELLITE IMAGES

Camilo Delgado-Correal & José E. García

**RESUMEN**

Se hace una revisión de los principios físicos involucrados en el procesamiento digital de imágenes satelitales, más específicamente en el tema de la calibración radiométrica de las mismas. Se muestra una descripción conceptual de los procesos relevantes de la interacción radiación-atmósfera y radiación-suelo con el objetivo de que el lector entienda con detalle que significa la información contenida en las imágenes satelitales.

*Palabras clave*: espectroradiometro, atenuación radiativa de la atmósfera, curva de reflectancia.

**ABSTRACT**

It does a revision about the physical principles involved in digital processing of satellite images, more specifically in radiometric calibration of them. It shows a conceptual description of the interaction between radiation and atmosphere and radiation and soil in order to help the reader understand in more detail which means the information contained in satellite images.

## INTRODUCCIÓN

El uso de imágenes satelitales por parte de nuestras entidades locales para satisfacer necesidades del país en temas como gestión ambiental, gestión de riesgo, sistemas productivos, recursos minerales y energéticos, planificación urbano regional, salud, seguridad y defensa, información básica, cambio climático y ordenamiento territorial (IGAC-CCE 2009) genera investigaciones que buscan cada vez tener una mejor calidad de las imágenes que se utilizan en dichas aplicaciones. Una de estas investigaciones se basa en corregir los defectos producidos por la atenuación de la radiación debido a su interacción con la atmosfera, que provoca una pérdida de nitidez en las imágenes capturadas por los respectivos sensores (Figura 1).

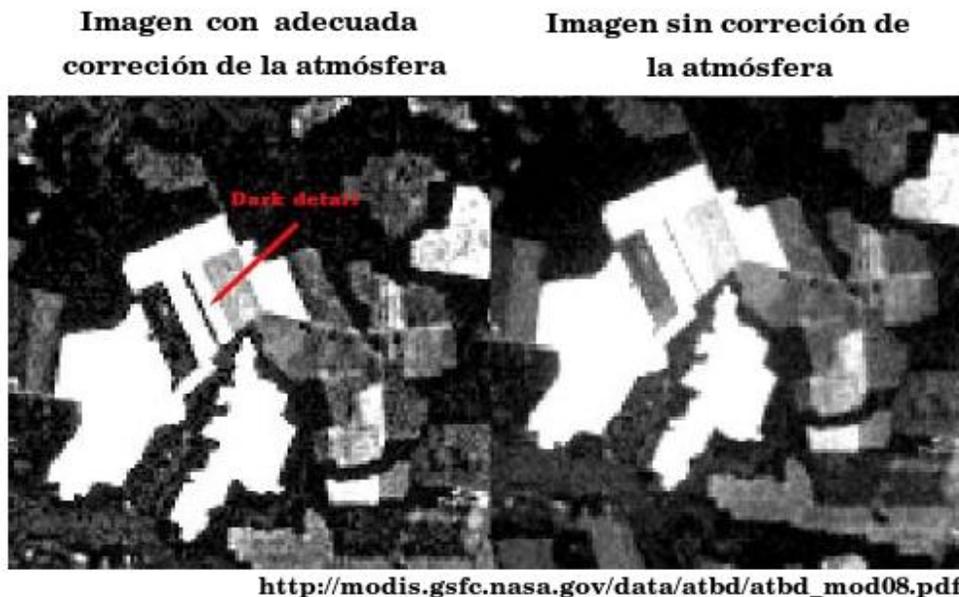

**Figura 1.** Diferencias entre una imagen con, y, sin corrección radiativa de la atmósfera.

En la actualidad existen rutinas al interior de los paquetes de procesamiento de imágenes satelitales que permiten corregir estos defectos, sin embargo la atmósfera en zonas tropicales no corresponde con los modelos de estos paquetes computacionales, cuyas constantes y parámetros corresponden a moldeamientos de latitudes altas en donde las condiciones atmosféricas son distintas. Esto genera que se haga necesario estudiar con detalle los conceptos físicos involucrados con el paso de la radiación electromagnética a través de la atmósfera y del suelo para luego utilizarlos en códigos computacionales que hallen la atenuación por longitud de onda de la atmósfera colombiana. Finalmente con la ayuda de un espectroradiómetro podremos ajustar la calibración radiométrica de las imágenes satelitales hechas con los respectivos códigos computacionales (Delgado-Correal & García 2011).

## ASPECTOS FÍSICOS DE LA TRANSFERENCIA RADIATIVA A TRAVÉS DE LA ATMÓSFERA

Para comenzar a hacer esta descripción observemos la siguiente figura,

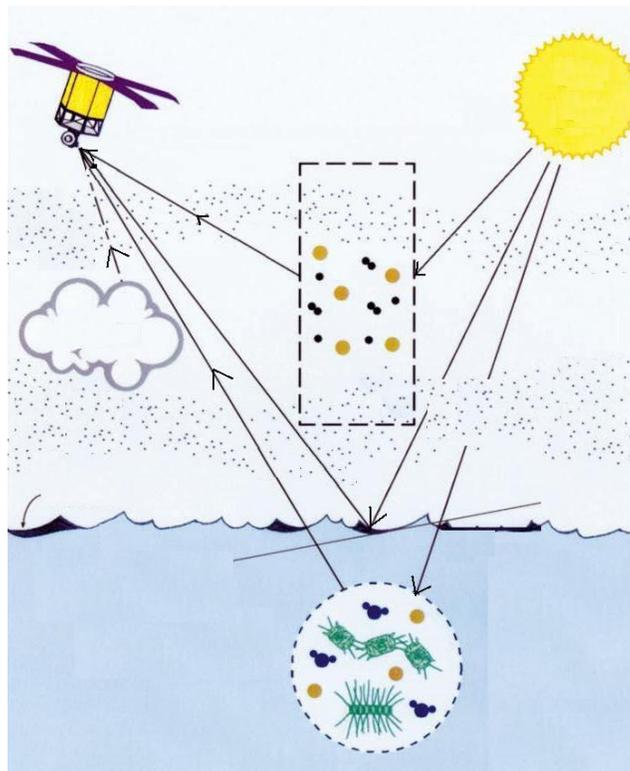

**Figura 2.** Esquema general del paso de la radiación de Sol a través de la atmósfera terrestre. Modificado de: (Hooker & McClain 2000).

En la figura 2 se puede observar que los sensores pasivos satelitales de observación de la Tierra reciben la información de la radiación proveniente de la superficie y de la atmósfera. En donde la radiación que proviene del sol, primero pasa por la atmósfera donde una parte es absorbida y otra es dispersada por los componentes de la atmósfera (principalmente compuesta de moléculas de $O_2$, $N_2$, $O_3$, $CO_2$ y de vapor de agua entre otros). Luego de su paso por la atmósfera llega e interactúa (dependiendo de la longitud de onda) con la superficie terrestre donde también una parte es absorbida y otra es dispersada. Finalmente la radiación que es reflejada por la superficie terrestre atraviesa la atmósfera de nuevo para llegar a ser detectada por los sensores satelitales de observación de la Tierra.

## *RADIACIÓN SOLAR*

La radiación solar es el insumo principal para obtener imágenes satelitales de la superficie terrestre, puesto que los sensores pasivos detectan la reflexión de la radiación emitida por los diferentes cuerpos que se encuentran sobre y debajo de la superficie terrestre. Por esta razón para comenzar a entender la información subyacente contenida en las imágenes satelitales, empezaremos haciendo la descripción física de la radiación proveniente del Sol. En la figura 3 se observa la irradiancia espectral solar en el tope de la atmósfera,

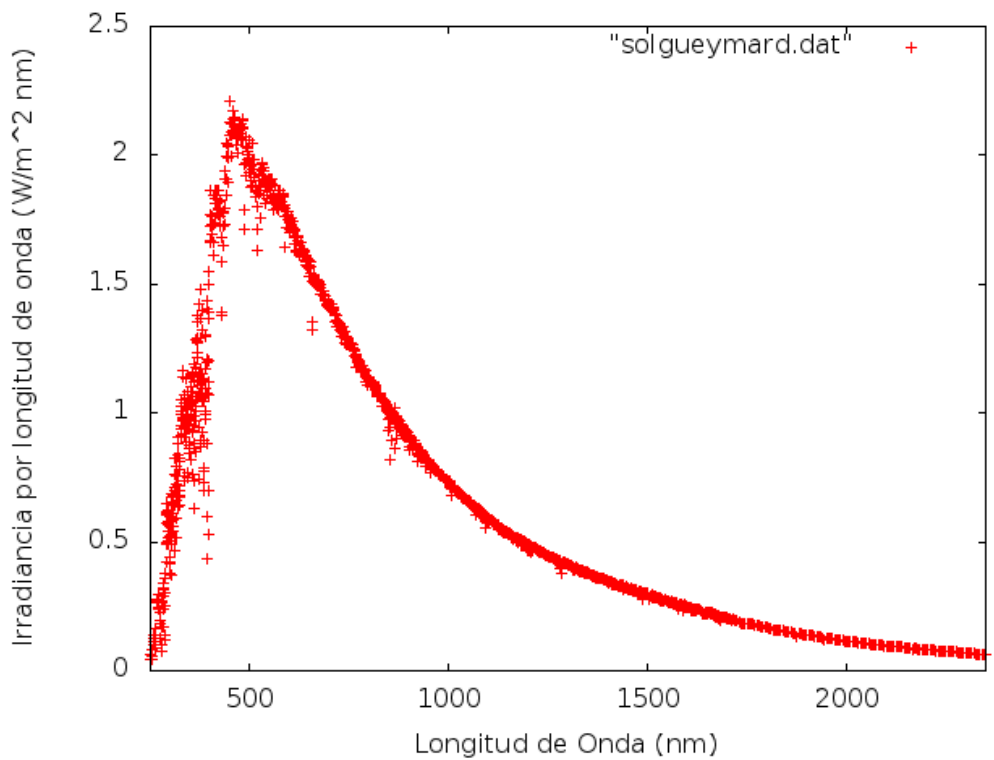

**Figura 3.** Espectro de la radiación solar en el tope de la atmósfera. (Adaptado de: Gueymard 2004).

La forma funcional de este espectro es muy similar a la función espectral de la emisión de radiación de un cuerpo negro (ver ajuste realizado en la figura 4), el cual es un hipotético cuerpo definido como un absorbente perfecto de toda la radiación incidente sobre él y que solamente radia en forma proporcional al aumento gradual de su temperatura a la cuarta potencia (Roy & Clarke 2003).

La distribución espectral de la radiación emitida por un cuerpo negro esta descrita matemáticamente por la siguiente relación (Planck 1901),

$$B_\lambda = \frac{2hc^2}{\lambda^5 \left( e^{\frac{hc}{\lambda K_B T}} - 1 \right)}$$

$Donde: h = $ Constante de Planck, $c = $ Velocidad de la Luz, (1)
$T = $ Temperatura absoluta, $\lambda = $ Longitud de onda, $K_B = $ Constante de Boltzmann

En la Ec. (1) podemos ver una intima relación entre la temperatura a la cual se encuentra el cuerpo negro y su irradiancia espectral.

Interesados en conocer la forma funcional de la irradiancia espectral solar suponemos que el Sol radia espectralmente como un cuerpo negro. Para mostrar esto debemos ajustar la Ec. (1) a los datos correspondientes a la irradiancia espectral solar mostrada en la Figura 3. Este ajuste se puede apreciar en la Figura 4.

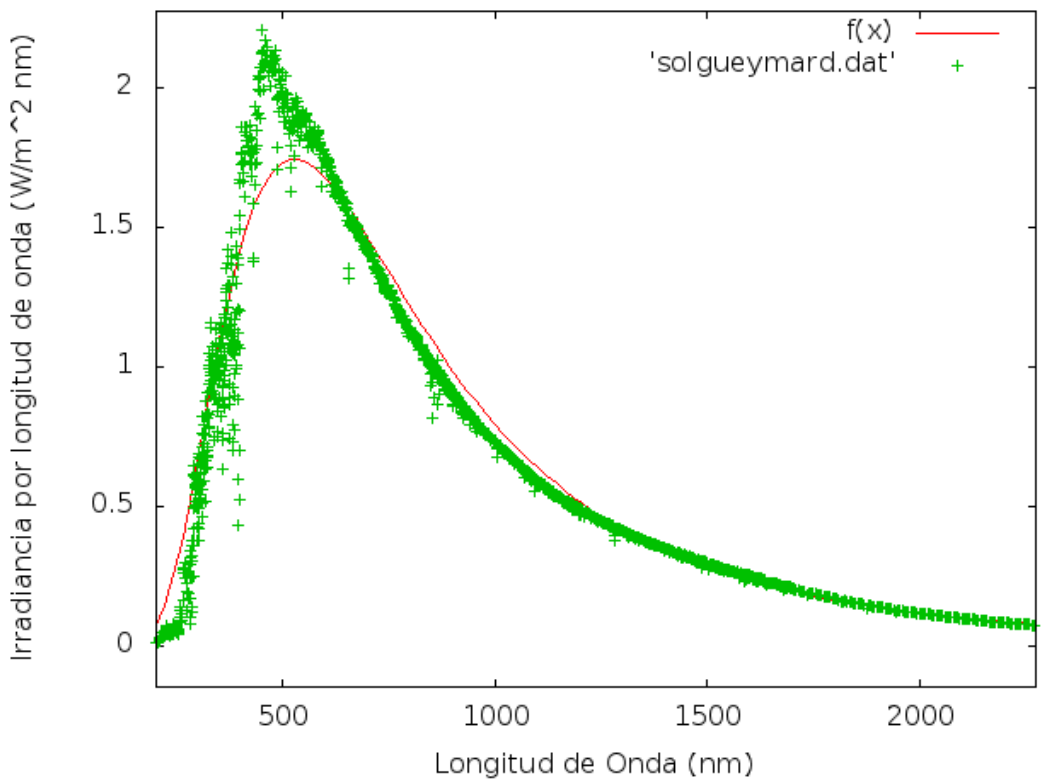

**Figura 4.** Ajuste de la función analítica de la emisión espectral de la radiación de un cuerpo negro a la irradiancia espectral solar.

En la Figura 4 se puede observar un buen ajuste entre la función analítica de la emisión espectral de la radiación de un cuerpo negro a la irradiancia espectral solar mostrando que se puede modelar la radiación del Sol como la radiación de un cuerpo negro. En este ajuste particular se encuentra una temperatura de color del Sol: $T_c$ = 5483.35 K, el cual es un dato muy cercano al valor de temperatura efectiva solar reportado por otros investigadores de 5778K (Fligge, M. et al. 1998) (http://nssdc.gsfc.nasa.gov/planetary/factsheet/sunfact.html).

A pesar de que la temperatura de color del Sol nos proporciona un indicativo de la temperatura superficial solar (Clayton 1968), la cantidad más importante para describir la temperatura superficial del Sol o de cualquier otra estrella es la temperatura efectiva $T_{eff}$ (Karttunen et al. 2007). Esta $T_{eff}$ es definida como la temperatura de un cuerpo negro, el cual radia con la misma densidad de flujo total que la estrella. Físicamente se puede considerar el Sol como un cuerpo negro puesto que para producir su radiación, en su interior se hacen continuamente reacciones de fusión nuclear (ver Figura 5) que tiene como resultado una producción autónoma de radiación.

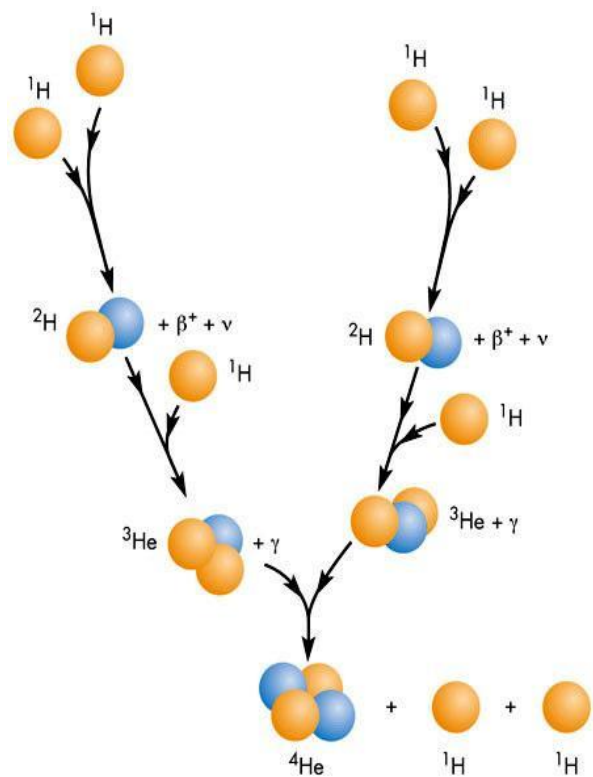

http://venus.ifca.unican.es/~carreraf/AstronomiaGeneral/Material/Sol.pdf

**Figura 5.** Esquema pictórico de las reacciones nucleares que ocurren en el interior del núcleo del Sol.

## PASO DE LA RADIACIÓN SOLAR POR LA ATMÓSFERA

Después que la radiación es emitida por el Sol y llega a la vecindad de la Tierra empieza a interactuar con las partículas que componen la atmósfera terrestre, las cuales son de tres tipos: moléculas de gas (de mayor a menor concentración: $N_2$, $O_2$, Ar, $CO_2$, Ne, He, $CH_4$, Kr, CO, $SO_2$, $H_2$, $O_3$, $N_2O$, Xe, $NO_2$, entre otras), partículas líquidas como los aerosoles, las nubes y, sólidas como la nieve. Estas partículas atenúan esta radiación provocando una disminución de su intensidad para el momento que llegue a la superficie terrestre. Este proceso de atenuación se divide en absorción y dispersión, y se puede observar en la figura 6 un ejemplo de ello.

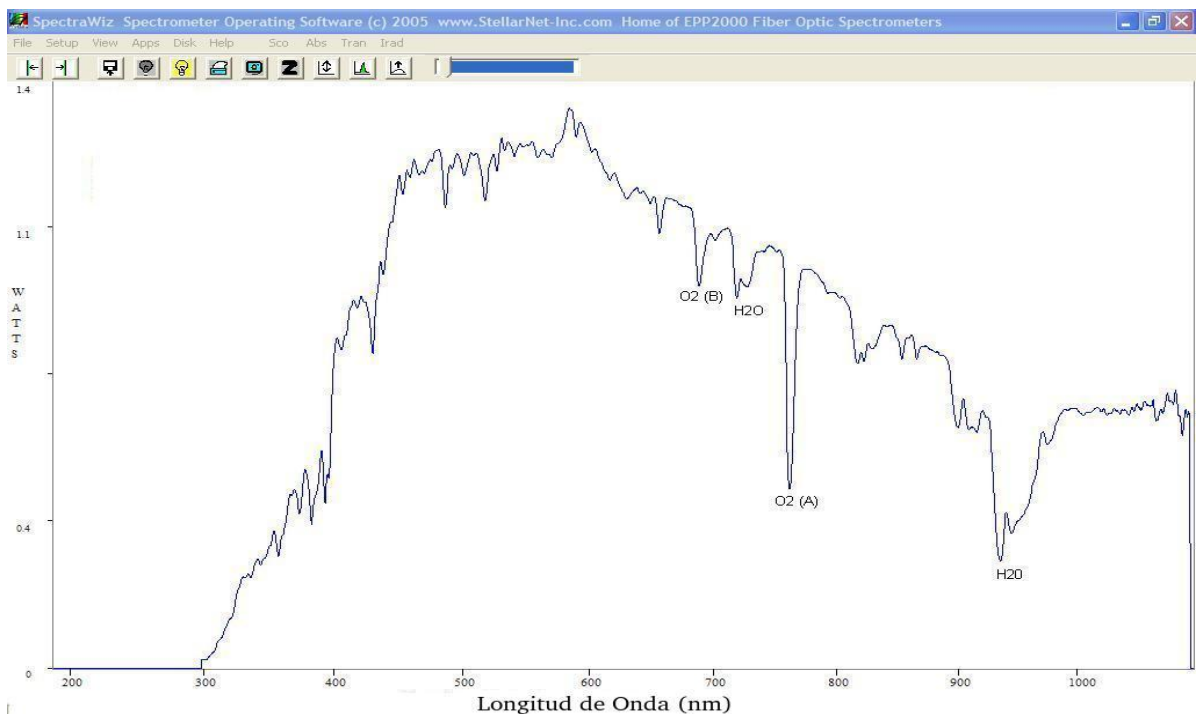

http://www.stellarnet-inc.com/images/BLACK-Comet%20Solar%20Spectrum_large.jpg

**Figura 6.** Espectro de la radiación del Sol en el visible e infrarrojo cercano tomado en la superficie de la Tierra con un espectroradiómetro.

La ecuación de transferencia radiativa que describe los procesos de absorción y dispersión de la radiación que pasa a través de un medio, en este caso de la atmósfera, está definida matemáticamente por la siguiente relación (Rees 2006),

$$\frac{dL_f(\theta,\varphi)}{dz} = -(\gamma_A)L_f(\theta,\varphi) + \gamma_s J_f(\theta,\varphi) \quad (2)$$

Donde $L_f$ es la radiancia[1] espectral que se propaga en el interior de la atmósfera, en una dirección arbitraria $(\theta,\varphi)$. $J_f$ es la radiación dispersada en la dirección $(\theta,\varphi)$ proveniente de otra dirección $(\theta',\varphi')$, dz corresponde un diferencial en la trayectoria de propagación de la radiación y $\gamma_A$ es el coeficiente de atenuación, que está definido como,

$$\gamma_A = \gamma_a + \gamma_s \quad (3)$$

Donde $\gamma_a$ y $\gamma_s$ son los coeficientes de absorción y de dispersión de la atmósfera respectivamente.

Estos coeficientes están íntimamente relacionados con la interacción a escalas cuánticas entre la radiación electromagnética que proviene del Sol y las moléculas presentes en la atmósfera. En esta concepción cuántica los electrones internos de los átomos que componen las moléculas presentan niveles de energía discretos. Al incidir un fotón[2] sobre un electrón, provoca que este aumente de nivel energía, es decir haya una absorción de la energía, que está definida por la siguiente relación (García & Ewert 2008),

$$\Delta E = h\,v = \frac{h\,c}{\lambda} \quad (4)$$

Hay tres mecanismos principales por los cuales las moléculas pueden absorber radiación: transiciones electrónicas, vibración y rotación. Un análisis detallado de cada una de ellas se encuentra explicado en: (Foot 2005) y (Atkins & Friedman 2005). Por eso dependiendo de la configuración electrónica de nuestra molécula vamos a tener una absorción de la radiación del Sol específica para determinado valor de longitud de onda para cada molécula (ver Tabla 1).

Observando con detalle la ecuación (2) es claro entender la importancia de hallar el coeficiente de atenuación de la atmósfera para poder describir teóricamente el proceso de transferencia radiativa a través de la misma. Para hallarla es muy importante describir con detalle los procesos de absorción y dispersión realizados por las moléculas presentes en la atmósfera, caracterizar la respuesta radiativa de las nubes[3], además de tener la información de la reflectancia del suelo para poder escribir con detalle la radiancia que proviene del mismo.

---

1 La radiancia espectral se define como la irradiancia por unidad de ángulo sólido, tal que sus unidades son (W/(sr·m$^2$)).

2 El fotón o cuanto de energía de la luz (light energy quantum) representa la cuantización del campo electromagnético. Este concepto fue introducido por Albert Einstein para dar una explicación al efecto fotoeléctrico (Einstein 1905).

3 Esta caracterización es muy importante en el caso del territorio colombiano donde el porcentaje de nubosidad es muy alto.

De otra forma al comparar la información radiométrica contenida en las imágenes capturadas con sensores satelitales con la información registrada por un espectroradiómetro de la radiación emitida por la superficie terrestre se puede hallar la intensidad total de la atención de la radiación del Sol en la atmósfera sobre una región específica de la Tierra.

**Tabla 1.** Principales valores de longitud de onda (líneas de absorción) en los cuales las moléculas que componen la atmósfera absorben radiación solar. Adaptados de: (Rees 2006)

| Longitud de onda (nm) | Molécula | Longitud de onda (nm) | Molécula |
|---|---|---|---|
| 260 | $O_3$ | 3900 | $N_2O$ |
| 600 | $O_3$ | 4300 | $CO_2$ |
| **690** | **$O_2$** | 4500 | $N_2O$ |
| **720** | **$H_2O$** | 4800 | $O_3$ |
| **760** | **$O_2$** | 4900 | $CO_2$ |
| 820 | $H_2O$ | 6000 | $H_2O$ |
| **930** | **$H_2O$** | 6600 | $H_2O$ |
| 1120 | $H_2O$ | 7700 | $N_2O$ |
| 1250 | $O_2$ | 7700 | $CH_4$ |
| 1370 | $H_2O$ | 9400 | $CO_2$ |
| 1850 | $H_2O$ | 9600 | $O_3$ |
| 1950 | $CO_2$ | 10400 | $CO_2$ |
| 2000 | $CO_2$ | 13700 | $O_3$ |
| 2100 | $CO_2$ | 14300 | $O_3$ |
| 2600 | $H_2O$ | 1500 | $O_3$ |

# MEDIDA IN-SITU DE LA REFLECTANCÍA DEL SUELO UTILIZANDO UN ESPECTRORADIÓMETRO

Lastimosamente la información que registran los sensores pasivos de observación de la Tierra no corresponde exactamente a la radiación emitida por el suelo, puesto que los gases constituyentes de la atmósfera dependiendo de la banda, absorben parte de su radiación. Por esta razón utilizar un espectroradiómetro para realizar calibraciones radiométricas *in-situ* de las imágenes satelitales se convierte en una gran herramienta para hacer tal fin (Delgado-Correal & García 2011), puesto que este dispositivo registra la radiación que proviene directamente de la superficie terrestre.

Específicamente este equipo instrumental permite colectar la radiación proveniente de la tierra, generalmente en las bandas del visible e infrarrojo cercano, presentando un registro continuo de la radiación emitida por la superficie en un ancho de banda espectral promedio entre los 300nm hasta los 1000nm, hay instrumentos que pueden ir hasta los 2500nm. Finalmente con estos datos de radiancía se puede estimar la reflectancía del suelo.

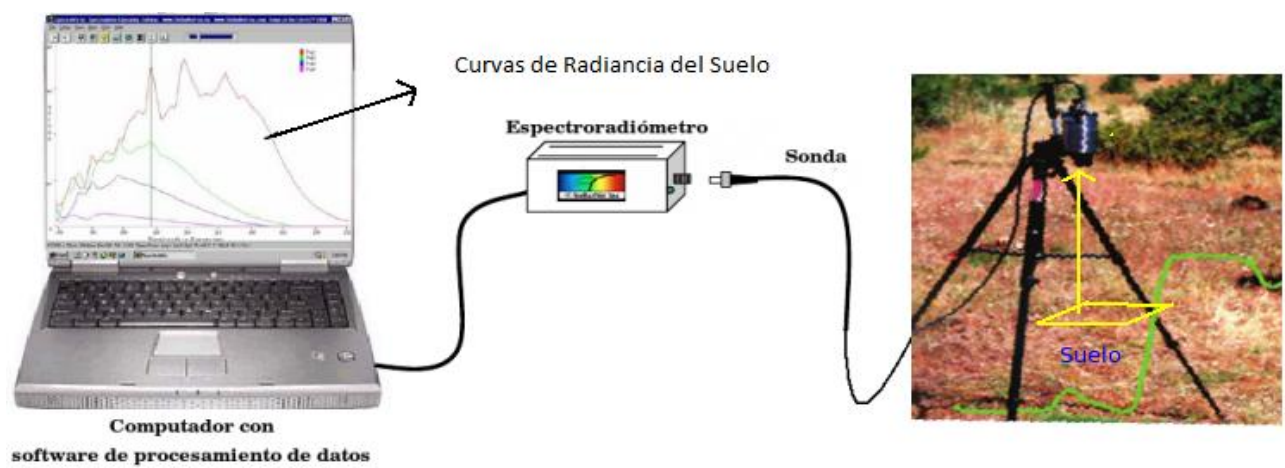

**Figura 7.** Esquema de las partes que componen la medición *in-situ* de la radiancía del suelo.

El ancho de escena por así decirlo de un espectroradiómetro es del orden de centímetros y para obtener la radiancía *un situ* de una región de 1 metro (resolución típica de las bandas espectrales de los sensores del satélite IKONOS) es necesario recorrer esta región y tomar las mediciones de la radiación (por longitud de onda) del suelo punto a punto a lo largo y ancho de nuestra región de interés y al final construir una grilla espacial con los valores obtenidos en cada medición. Esta medición debe ser realizada en el menor tiempo posible para considerar que en todas las mediciones tenemos las mismas condiciones de iluminación solar y atmosférica.

Es importante mencionar que las mediciones de reflectancía del suelo colombiano han sido hasta ahora ejercicios aislados de investigación universitaria y de centros de investigación, sin que el país cuente ni con los equipos necesarios, ni con los programas de investigación adecuados para hacer el mapeo total de la reflectancía de la superficie colombiana con el objetivo principal de poder definir técnicamente las bandas espectrales con las cuales deben contar los sensores de percepción remota de un futuro satélite colombiano de observación de la Tierra que pueda suplir la mayoría de las necesidades colombianas en el tema de percepción remota.

## *INTERACCIÓN FÍSICA DE LA RADIACIÓN SOLAR CON LA SUPERFICIE TERRESTRE*

Después de pasar por la atmósfera la radiación proveniente del Sol interactúa con la superficie terrestre, donde parte de ella es absorbida, transmitida, dispersada y reflejada (Figura 8) cumpliendo con la siguiente relación,

$$E_I = E_R + E_A + E_T \rightarrow E_R = E_I - (E_A + E_T)$$
$$E_I = \text{Energía incidente}$$
$$E_R = \text{Energía reflejada} \quad (5)$$
$$E_T = \text{Energía transmitida}$$

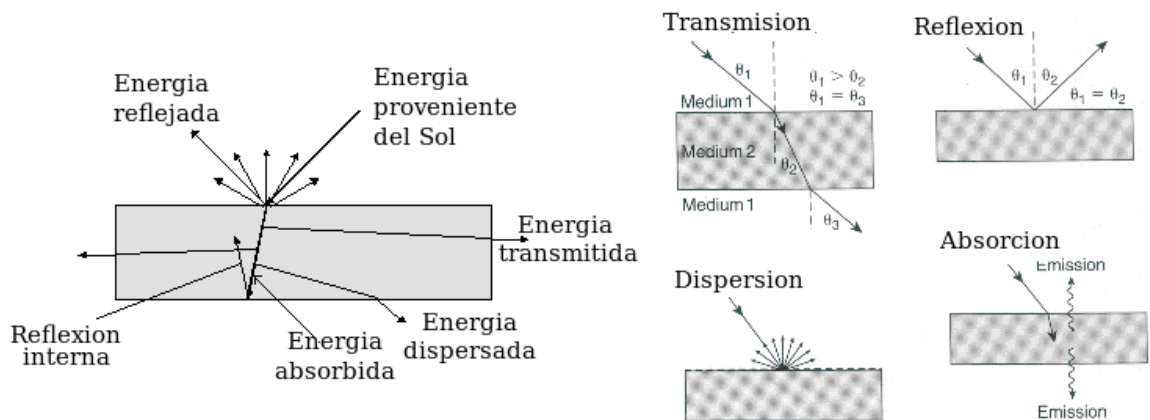

Modificado de: http://www.slrss.cn/rsgisforum/bbs/viewthread.php?tid=2372

**Figura 8**. Procesos macroscópicos de la interacción de la radiación del Sol, que atravesó la atmósfera, con la superficie terrestre.

La radiación reflejada por la superficie es mucho menor que la emitida por el Sol (Figura 9), esto debido a que gran parte es absorbida, dispersada y reflejada por los componentes de la atmósfera y del suelo.

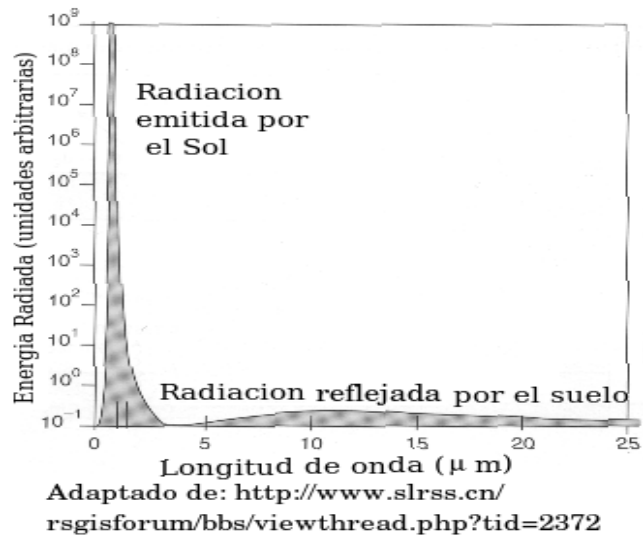

**Figura 9.** Comparación entre la radiación reflejada por el suelo y la emitida por el Sol.

La radiancía de la superficie de la Tierra, la cual está íntimamente ligada con los componentes del suelo, puede ser cuantificada mediante la medición de la porción de la radiación del Sol incidente que es reflejada. Esta medición es conocida como reflectancía espectral y muestra la respuesta radiativa de absorción, dispersión y de transmisión de los cuerpos que se encuentran debajo y sobre la superficie.

Un ejemplo típico de una curva de reflectancía real se muestra en la Figura 10 y fue construida aplicando la ecuación (6) a los datos obtenidos por un espectroradiómetro de la radiación del Sol y de la radiación emitida por 4 diferentes especies vegetales, bajo las mismas condiciones de iluminación.

$$\rho_\lambda = \frac{E_R(\lambda)}{E_I(\lambda)} * 100 \quad (6)$$

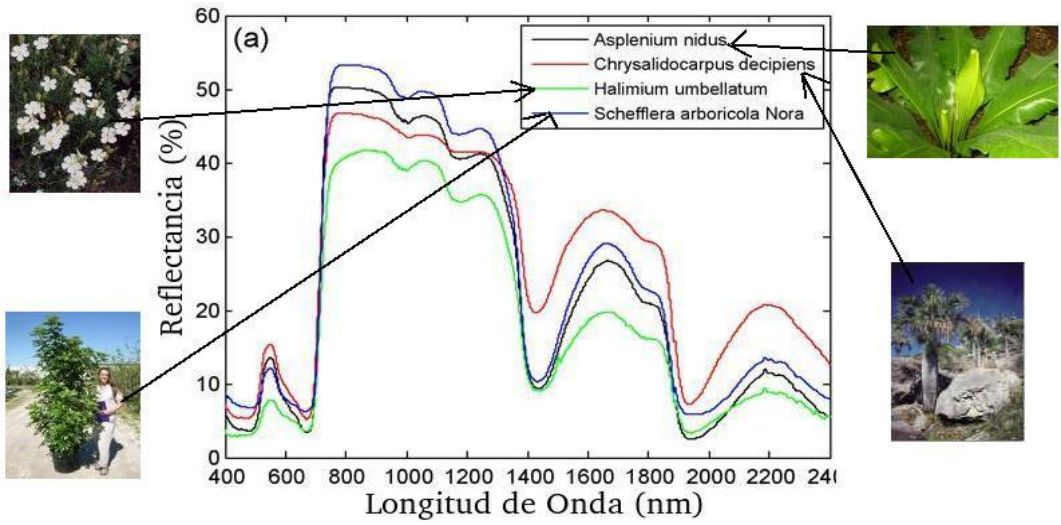

**Figura 10.** Curvas de reflectancía espectral de diferentes especies vegetales. (Modificado de: (Darvishzadeh & et al. 2008 ).

Además de su dependencia con la composición del suelo, la reflectancía del mismo depende de las condiciones de iluminación al cual este sometido, es decir de la posición a la cual se encuentren el Sol y la Luna en el transcurso del tiempo. Así mismo la información contenida en las imágenes que son registradas por un sensor satelital de observación de la superficie va a depender de los ángulos que relacionan la posición del Sol con la posición del sensor.

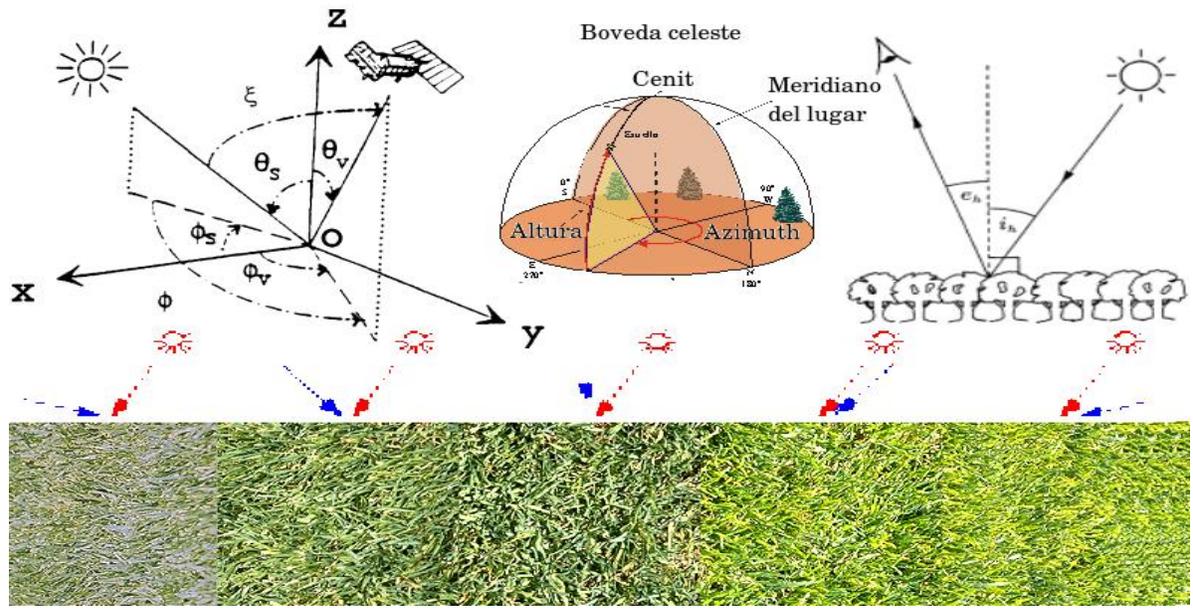

**Figura 11.** Dependencia del valor de la radiancía del suelo con la inclinación del sensor. Adaptado de: (Ducheim 1999), (Dymond 20001) y (Guiz 2010).

Por esta razón es necesario conocer la trayectoria de paso de los satélites y la posición del Sol sobre nuestra bóveda celeste, es decir conocer sus respectivos ángulos de altura y de azimut, para así poder utilizar las mediciones *in situ* de reflectancía de un terreno para calibrar radiométricamente nuestras imágenes satelitales.

Para explicar con detalle el fenómeno mostrado en la figura 11, consideremos para comenzar el siguiente esquema,

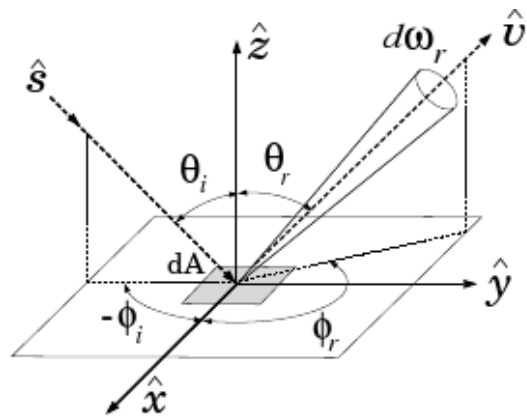

**Figura 12.** Esquema geométrico de las direcciones de la irradiancía solar y de la radiancía del suelo. Fuente: (Darvishzadeh & et al. 2008 ).

En la figura 12 podemos observar a un elemento superficial dA, que es iluminado por una fuente de radiación, como el Sol, que se encuentra en una dirección arbitraria $\hat{s}=(\theta_i,\varphi_i)$, y la radiación emitida por el suelo es capturada por un sensor[4] que está en una dirección $\hat{s}=(\theta_r,\varphi_r)$. Los $(\theta_i,\varphi_i)$ ángulos corresponden respectivamente al ángulo polar (90° - altura) y al azimut; así mismo $d\omega_r$ es el ángulo solido subtendido por el sensor desde cualquier punto de la superficie. De esta manera siguiendo la notación anterior podemos definir la irradiancía, el cual es el flujo de radiación incidente sobre la superficie mediante la siguiente relación,

$$E(\theta_i,\varphi_i)=\frac{d\Phi(\theta_i,\varphi_i)}{dA} \quad (7)$$

---

4  Que puede estar colocado en una plataforma satelital o puede estar tan solo unos metros sobre la superficie, como un espectroradiómetro.

El brillo de la superficie registrado por el sensor es proporcional a la radiancía del suelo, el cual se propaga en una trayectoria definida por los ángulos $(\theta_r, \varphi_r)$ y es definida por,

$$L_r(\theta_r, \varphi_r; \theta_i, \varphi_i) = \frac{d\Phi(\theta_r, \varphi_r; \theta_i, \varphi_i)}{dA\cos(\theta_r)d\omega_r} \quad (8)$$

Donde podemos ver que la imagen de la superficie terrestre capturada por un sensor depende explícitamente de la posición de la fuente de iluminación y del sensor.

La relación entre la irradiancía del Sol y la radiancía de una superficie, es una medida cuantitativa de la reflectancía del suelo, y es conocida como *Bi-directional reflectance distribution function* (BRDF) y definida mediante,

$$BRDF = \frac{dL_r(\theta_r, \varphi_r; \theta_i, \varphi_i)}{dE(\theta_i, \varphi_i)} \quad (9)$$

Comparando la ecuación (8) con (6) podemos ver que ambas ecuaciones son semejantes, donde (6) es muy útil para obtener curvas de reflectancía del suelo y (8) sirve para entender el significado físico detrás de estas curvas.

**CONCLUSIONES**

En este trabajo pudimos observar que tanto la atmósfera como el suelo generan parámetros relacionados con la atenuación de la radiación electromagnética proveniente del Sol que va a quedar registrada en los sensores de percepción remota provocando que la calidad de nuestras imágenes satelitales se vea afectada por ella. Sin embargo con una adecuada inclusión de estos procesos físicos en las rutinas de calibración radiométrica podremos obtener imágenes que correspondan más a la realidad.

**AGRADECIMIENTOS**

desarrollo satelital y aplicaciones en el tema de observación de la tierra. Los autores agradecen a los investigadores Iván López, Andrés Franco y Germán Gaviría por su participación de las continuas discusiones realizadas a lo largo de este trabajo.

**REFERENCIAS BIBLIOGRÁFICAS**

CAMILO DELGADO-CORREAL

Físico, Universidad Nacional de Colombia, Magister en Ciencias-Astronomía, Observatorio Astronómico Nacional, Profesor de vinculación especial Universidad Distrital Francisco José de Caldas, Integrante del semillero GEIPER adscrito al Núcleo de Investigación en Datos Espaciales (NIDE).
Correo electrónico: mcdelgadoc@unal.edu.co

JOSE E. GARCIA

Físico, Universidad Nacional de Colombia, Especialización en administración tecnológica y de innovación, Universidad Simon Fraser, Canadá. Director del grupo de Física Aplicada y Desarrollo tecnológico, Centro Internacional de Física.
Correo electrónico: jose.garcia@cif.org.co